\documentclass[twocolumn,10pt]{IEEEtran}
\usepackage{ifpdf, flushend,subfigure}

\ifCLASSINFOpdf
  \usepackage[pdftex]{graphicx}
  \graphicspath{{../pdf/}{../jpeg/}}
  \DeclareGraphicsExtensions{.pdf,.jpeg,.png}
\else
  \usepackage[dvips]{graphicx}
  \graphicspath{{../eps/}}
  \DeclareGraphicsExtensions{.eps}
\fi
\usepackage{graphicx}
\usepackage{epstopdf}
\graphicspath{{../}}
\usepackage{multirow}
\usepackage{float}
\usepackage[cmex10]{amsmath}
\usepackage {amssymb}
\usepackage{array}
\usepackage{mdwmath}
\usepackage{mdwtab}
\usepackage{eqparbox}
\usepackage{url}
\usepackage{nomencl}
\usepackage{cite}
\usepackage{algorithm}
\usepackage{algorithmic}
\usepackage{makeidx}
\usepackage{ifthen}
\usepackage{subfigure}
\usepackage{caption}
\usepackage{pdflscape}
\usepackage{hyperref}
\usepackage{xcolor}
\newcommand{\beq}{\begin{equation}}
\newcommand{\eeq}{\end{equation}}
\newcommand{\beqn}{\begin{eqnarray}}
\newcommand{\eeqn}{\end{eqnarray}}
\newcommand{\beqno}{\begin{eqnarray*}}
\newcommand{\eeqno}{\end{eqnarray*}}
\newcommand{\bma}{\begin{displaymath}}
\newcommand{\ema}{\end{displaymath}}
\newcommand{\bnu}{\begin{enumerate}}
\newcommand{\enu}{\end{enumerate}}
\newcommand{\bce}{\begin{center}}
\newcommand{\ece}{\end{center}}
\newcommand{\btb}{\begin{tabular}}
\newcommand{\etb}{\end{tabular}}

\usepackage{ragged2e}

\begin{document}
% Start cover letter

\newpage

% \title{Uncertain Renewable Energy and EVs Integrated with Network Expansion Planning Considering Dynamic Thermal Line Rating}
\title{Optimal Network Expansion Planning With    Renewable Energy and Electric Vehicle Integration Considering Dynamic Thermal Line Rating}

\title{Optimal Network Expansion Planning  Considering Uncertain  Dynamic Thermal Line Rating and Renewable Energy Generation}

\title{Optimal Network Expansion Planning  Considering Uncertain  Dynamic Thermal Line Rating}
\author{\IEEEauthorblockN{Arash~Baharvandi, and~Duong~Tung~Nguyen,~\IEEEmembership{Member,~IEEE}}  %\vspace{-0.2cm}
\thanks{The authors are with the School of Electrical, Computer and Energy Engineering, Arizona State University, Tempe, AZ, United States. Email: \textit\{abaharv1,~duongnt\}@asu.edu.
 %(\textit{Corresponding author}: Duong Tung Nguyen).
 \\This work has been submitted to the IEEE for possible publication. Copyright may be transferred without notice, after which this version may no longer be accessible.}}


\maketitle

\begin{abstract}
This paper examines the integrated generation and transmission expansion planning problem to address the growing challenges associated with increasing power network loads.  The proposed approach optimizes the operation and investment costs for new generation units and transmission lines, while also considering the environmental benefits of integrating renewable energy sources (RES) and the impact of electric vehicle (EV) charging on the grid. % traveling between cities.
%To effectively manage the inherent uncertainties in demand, EV charging loads, and RES generation, a hybrid approach that combines stochastic and robust optimization is employed. 
The inherent uncertainties in demand, EV charging loads, and RES generation are managed using a hybrid stochastic-robust optimization approach. % method combining robust optimization and stochastic programming. 
 %To manage uncertainties in demand, EV charging loads, and RES outputs, the proposed methodology utilizes a hybrid robust-stochastic programming approach that combines robust optimization with stochastic programming.
 %Furthermore, we incorporate Dynamic Thermal Line Rating (DTLR) into the expansion planning model to enhance transmission line efficiency and resiliency. 
 Additionally, the model integrates Dynamic Thermal Line Rating (DTLR) to improve the efficiency and resilience of transmission lines. The framework also tackles the uncertainty related to DTLR, incorporating a heuristic linearization technique to reduce model complexity. The effectiveness of the proposed model and techniques is evaluated through simulations conducted on two case studies: the modified IEEE 6-bus system and the IEEE 24-bus Reliability Test System.

\end{abstract}
%\vspace{-0.15in}
\begin{IEEEkeywords}
Integrated generation and transmission expansion planning, uncertainty, dynamic thermal line rating. %renewable energy resources, electric vehicles, dynamic thermal line rating.
\end{IEEEkeywords}
%\IEEEpeerreviewmaketitle
%\vspace{3pt}

\printnomenclature

% \section{abstract material}

\section{Introduction}

As energy consumption continues to rise, it is crucial to determine
%incorporate the costs of installing new equipment and units into the planning process. Determining 
 the type, capacity, and timing of new generation units and transmission lines for integration into the network to minimize overall investment and operational costs while meeting technical and operational constraints \cite{stoll1989least}. This approach ensures that power infrastructure expansion remains economically viable. % \cite{kim2015integrated}. 
 Renewable energy sources (RESs) offer a promising solution to reduce greenhouse gas emissions and operational costs, despite the uncertainty in their power output \cite {hatziargyriou2007microgrids}.
%despite the inherent uncertainty in their output, to reduce greenhouse gas emissions and operational costs. 
The expansion planning process must also account for the increasing adoption of electric vehicles (EVs) and the impact of charging loads on the power grid \cite{luo2023coordinative}. 
%Traditional transportation systems that rely on fossil fuels have significant negative environmental impacts. In contrast, electrification of transportation presents a promising and more sustainable alternative \cite{luo2023coordinative}. 
%Additionally, Dynamic Thermal Line Rating (DTLR)  improve the efficiency and resilience of transmission lines.

%In some researches, generation and transmission expansion planning has been considered separately.
Previous research often considers generation and transmission expansion planning separately.
For example, generation expansion planning (GEP) is examined in \cite{dai2021incorporating,choi2022genetic,diewvilai2024enhancing}. Reference \cite{dai2021incorporating} integrates GEP with an investigation into external network flexibility. %and external flexibility is considered as a feasible set. 
In \cite {choi2022genetic}, the authors apply Genetic Algorithms to address the complexities of nonlinear GEP problems with reactive power planning. %A non-linear dynamic problem such as GEP with reactive power planning is challenging to solve directly due to its complexity. Consequently, methods like Genetic Algorithms can be effectively applied to address these challenges, as demonstrated in  \cite {choi2022genetic}. 
In \cite{diewvilai2024enhancing}, GEP is examined with varying demands throughout the year, incorporating flexibility to account for load changes and intermittent RESs.

% A GEP problem considering different demands within a year is represented in \cite{diewvilai2024enhancing}, and an appropriate flexibility has been presented to take change of load and intermittent RESs into account. 

Transmission expansion planning (TEP) is explored in \cite{yuan2022resilience,esmaili2020transmission,chen2023climate}. 
For instance, \cite{yuan2022resilience} proposes a TEP problem considering network configuration using switching to enhance resilience during typhoons. 
%A TEP problem considering network configuration using switching is proposed in \cite{yuan2022resilience} to tackle a typhoon in a resilient network. 
In \cite{esmaili2020transmission}, thyristor-controlled series compensators (TCSCs) and superconducting fault current limiters (SFCLs) are used in a TEP problem 
to increase the maximum allowable power moving through the lines and restrain short-circuit currents, respectively. 
A robust TEP model, incorporating a climate-adaptive uncertainty set (CUS), is developed in 
\cite{chen2023climate} to achieve enhanced security in operations amidst climate change while minimizing line investment costs. %\cite{chen2023climate} introduces a climate-adaptive Transmission Network Expansion Planning (TNEP). This approach essentially involves a robust TNEP that incorporates a climate-adaptive uncertainty set (CUS). The process for determining the CUS involves three key steps.

There is a growing literature on integrated generation and transmission expansion planning (IGTEP). In \cite{amini2023multi}, the authors explore a multi-objective IGTEP problem, focusing on voltage stability and power losses. Reference \cite{gonzalez2018generation} studies an IGTEP problem involving RES integration, where the RESs are shared between two AC-independent systems connected by HVDC grids. In \cite{aghaei2014generation}, an IGTEP model is proposed that incorporates reliability criteria, using forced outage rates for lines and units, and includes an expected-energy-not-supplied term in the objective function to address reliability concerns.

Recent research has also investigated various IGTEP problems under uncertainty. % \cite{munoz2021integrated,yin2020generation,garcia2023computational,rintamaki2023achieving}. %Integrated generation and transmission expansion planning (IGTEP) problem with its uncertain nature is considered in . 
%GEP and TEP are optimized independently resulted in suboptimal solutions, but 
For example, \cite{munoz2021integrated} presents an IGTEP model that addresses RES uncertainties using stochastic programming (SP).
%IGTEP is solved in the presence of RESs. Its uncertainties are addressed by stochastic programming (SP) through different scenarios. 
In \cite{yin2020generation}, a decomposition method is applied to IGTEP to manage intractability, incorporating robust optimization (RO) and SP to handle contingencies and load uncertainties.
%while the uncertainties of element contingency and load are implemented by robust optimization (RO) and SP. 
%This method follows column-and-constraint generation and L-shaped algorithms. 
A two-stage stochastic model for IGTEP is proposed in \cite{garcia2023computational}, utilizing an accelerated solving approach and evaluating value-at-risk. In \cite{rintamaki2023achieving}, stochastic adaptive robust optimization is employed to identify optimal multi-year investment strategies that reduce greenhouse gas emissions while accounting for uncertain demand and renewable energy generation.

%\vspace{0.6in}
Traditional network planning often relies on static thermal ratings, which can lead to conservative designs and underutilized capacities. Dynamic Thermal Line Rating (DTLR) offers a significant improvement by providing system operators with real-time insights into the power transfer capabilities of transmission lines as they fluctuate with changing weather conditions \cite{maksic2016dynamic}. In this work, we explore the integration of DTLR into the planning process, where line ratings are adjusted based on real-time environmental conditions. This approach enhances resource utilization and can result in substantial cost savings.
Recent studies have examined expansion planning with DTLR. %\cite{li2023active,numan2020impact,jabarnejad2021linearized}]. 
Reference \cite{li2023active} addresses expansion planning in distribution networks, considering dynamic thermal ratings for lines and transformers, suggesting that underground cables are more suitable for urban areas. 
In \cite{numan2020impact}, the authors show that IGTEP incorporating DTLR and optimal switching is more cost-effective. 
% IGTEP problem considering DTLR and optimal switching is observed in \cite{numan2020impact} to be more effective in terms of investment cost. 
In \cite{jabarnejad2021linearized}, DTLR and optimal switching are integrated into a nonlinear IGTEP model, which is then linearized into a mixed-integer linear program (MILP).

This paper addresses generation and transmission expansion planning considering the presence of wind turbines, PVs, and %a fixed number of
EVs. % throughout the planning horizon. 
Given the inherent uncertainties associated with RESs, EV charging, % through the transmission network, 
and demand, we employ a method combining RO and SP to tackle these challenges. %The probability distribution function (PDF) for the uncertain parameters is assumed to follow a normal distribution, and we also discuss how the proposed approach can handle other distributions. 
Additionally, DTLR is integrated to enhance line utilization and resiliency, capturing the effects of seasonal weather changes on line capacity within the IGTEP framework. 
%When incorporating DTLR, 
The heat balance equation (HBE) is used to manage power limitations on the lines by substituting the power transfer constraint with
%When incorporating DTLR, the heat balance equation (HBE) is used to manage power limitations on the lines. Specifically, the constraint related to power transfer through the lines is replaced by 
a constraint on the maximum allowable temperature of the lines. The uncertainty introduced by variable weather conditions, impacting DTLR, is addressed using the proposed hybrid stochastic-robust optimization approach. We introduce a novel linearization method that significantly reduces computational complexity.  %is managed through the method described. 
% All in all, in none of the previous researches, uncertainty is not implemented while benefiting from both RO and SP, and unique linearization in this paper can reduce computations remarkably.
Our contributions can be summarized as follows:
\begin{itemize}
  \item \textit{Modeling}: We present an innovative IGTEP model that integrates DTLR and effectively captures the uncertainties in renewable energy generation, demand, and the variability of DTLR caused by changing weather conditions. 
  
  \item \textit{Solution approach: }  We employ a hybrid strategy combining RO and SP to address the IGTEP problem under uncertainty. We also propose an enhanced linearized AC load flow model and introduce an effective linearization technique to handle the IGTEP problem, particularly the complexities arising from 
  the DTLR equations.
 % \item Applying a unique linearization in the IGTEP problem and DTLR equations.
  \item \textit{Numerical results: } The proposed approach is validated through case studies on the modified IEEE 6-bus system and the IEEE 24-bus Reliability Test System, demonstrating the benefits of DTLR in enhancing grid flexibility and resilience. Additionally, we compare the computational times between the linear and non-linear formulations. % in terms of reduction in computational time. 
\end{itemize}

% \duong{Traditional network planning often relies on static thermal ratings, which can lead to conservative designs and underutilized capacities. This research explores the integration of DTLR, which adjusts the line ratings based on real-time weather and environmental conditions, into the planning process.  Our findings suggest that incorporating DTLR in network planning can lead to more efficient resource utilization and significant cost savings. The proposed approach is validated using a case study on a representative power system, demonstrating the benefits of adaptive line ratings in enhancing grid flexibility and resilience.}

The remainder of this paper is organized as follows: Section \ref{system model} describes the system model. Section \ref{sec:formu} presents the problem formulation. 
%The probability density functions of uncertain parameters are discussed in Section \ref{uncertainty}, followed by 
The linearization technique is introduced in Section \ref{Linearization}. Section \ref{Results}  provides the simulation results, and Section \ref{Conclusion} concludes the paper.

% The rest of this paper is organized as follows. Section \ref{system model} describes the system model. The problem formulation is presented in Section \ref{sec:formu}. Section \ref{uncertainty} discusses the probability density functions of uncertain parameters, followed by the linearization technique in Section \ref{Linearization}. Section  \ref{Results} presents the simulation results. Finally, the conclusions are presented in Section \ref{Conclusion}.
\begin{figure}[h!]
	\centering		\includegraphics[width=0.45\textwidth,height=0.15\textheight]{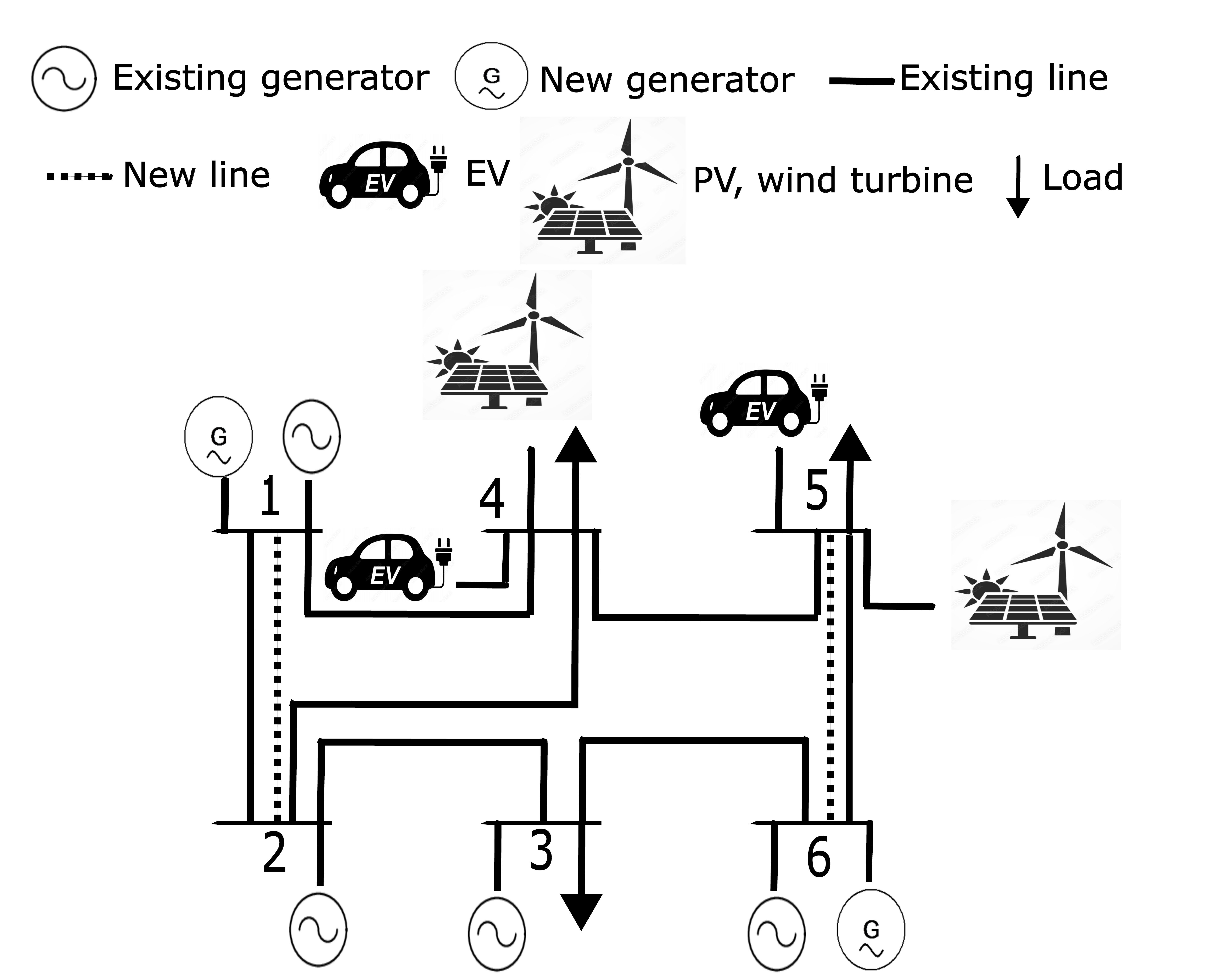}
			\caption{System Model}
	\label{fig:model}
\end{figure}
\vspace{-0.5cm}
\section{System Model}
\label{system model}

This paper addresses a planning problem focused on generation and transmission expansion in power systems. The proposed model is designed to tackle the challenges of integrating RESs and EVs, as illustrated in Figure \ref{fig:model}. 
The problem is approached from the perspective of a system operator responsible for optimizing %generation and transmission
network infrastructure to meet future demand and operational requirements. 
The operator's goal is to determine the optimal strategy for expanding both generation and transmission resources.
%
% The proposed model also integrates DTLR, a technology that adjusts the thermal rating of transmission lines based on real-time temperature measurements. In practice, various factors such as ambient temperature, wind speed, and other environmental factors can dynamically affect the capacities of transmission lines. Incorporating DTLR ensures that the transmission network can handle varying loads safely under different weather conditions. It also helps optimize line utilization and maintain resilience while adapting to seasonal and weather-induced variations.
% Our proposed model aims to support the integration of renewable energy sources and EVs into the power grid, addressing the challenges posed by their uncertainty and unique operational characteristics.
% Specifically, the goal of the operator is to minimize the total investment and operational costs while respecting all the operational constraints such as supply-demand constraints, and line and generator capacity constraints. Incorporating these constraints into the planning model helps in designing a power system that not only meets current and future demand but also operates within safe and efficient limits. Balancing these constraints effectively is crucial for maintaining the stability and resilience of the power network.
%%
%
%
A key feature of the model is the integration of DTLR, which dynamically adjusts thermal ratings of transmission lines based on real-time environmental conditions, such as wind speed and ambient temperature. 

By accounting for these factors, DTLR enhances 
transmission line capacity, ensuring safe operation under varying weather conditions.  Incorporating DTLR allows the model to optimize line utilization, enhance system resilience, and adapt to seasonal and weather-induced fluctuations.
The model is specifically designed to facilitate the integration of RES and EVs into the power grid, addressing the uncertainties and operational complexities they introduce. The primary goal is to minimize total investment and operational costs while adhering to all operational constraints, including supply-demand balance and line and generator capacity limits. These constraints are essential to ensuring that the system meets current and future demands while operating safely and efficiently.

Let $\Upsilon_{L}$ represent the set of all lines, including both existing and candidate new lines. The set of all generators is denoted by $\Upsilon_{G}$. We define
 $\Upsilon_{L}^{\sf  N}$ and $\Upsilon_{G}^{\sf N}$ as the set of candidate new lines and the set of candidate new generators. Let $c$, $g$, and $d$ signify the line, generator, and period indexes, respectively. 
% Additionally, period $d$ is part of the set $P_d$. 
The binary variable $u_c$ indicates whether the operator installs a new line $c$ and the binary variable $u_g$ indicates if a new generator $g$ is installed. 
%Two binary variables are defined to represent the status of the new lines and generators, denoted by $u_c$ and $u_g$. 
Additionally, let $P_{g,d}$ represent the output power of generator $g$ in period $d$. 
We define $\Delta_d$ as the duration of period $d$. 
The installation cost for a new line $c$ is $NC_c$ and the installation cost for a new generator $g$ is $NC_g$. The operational cost of generator $g$ is $CO_g$.  % is based on hour multiplied by 
We define  $P^{\sf max}_g$ as the maximum power generation capacity of generator $g$. The forecast base load, wind, and PV power generation are denoted by $P^{P}_{l,d}$, $wp^{P}_{l,d}$, and $pv^{P}_{l,d}$, respectively. 
The forecast EV charging demand at the bus connected to load $l$ in period $d$ is $EV^{P}_{l,d}$.
The line parameters—conductance, susceptance, and resistance— are indicated by $G_c$, $\beta_c$, and $R_c$, respectively. 

\begin{table}[ht] 
\centering
\caption{Notations}
\begin{tabular}{|l|l|}
\hline
Notation   & Meaning\\
\hline	
\multicolumn{2}{|c|}{\textbf{Set and indices}}\\
\hline	
$\Upsilon_{L}^{\sf  N}, \Upsilon_{G}^{\sf N}$ & Set of candidate new lines/generators\\
\hline	
% $\Upsilon_{G}^{\sf N}$ & Set of new units added to network\\
% \hline	
$\Upsilon_{L}, \Upsilon_{G}$ & Set of all lines/generators\\
\hline	
$c, g, d$   & Line index, generator index, and period index \\
\hline 
% $d$ & Time index \\
% \hline
$\mho_b$ & Set of buses\\
\hline	
$\mho^{ g}_b$ & Set of new and existing generators at bus $b$\\

\hline
$s(c), r(c)$ & Sending bus and receiving bus of line $c$\\
\hline
% $r(c)$ & Receiving bus of line $c$\\
% \hline	
\multicolumn{2}{|c|}{\textbf{Parameters}}\\
\hline
$NC_c$ & Cost of installing new line $c$\\% (\$) \\
\hline
$NC_g$ & Cost of installing new generator $g$ \\%(\$) \\
\hline
$\Delta_d$ & Time of operation related to period $d$ \\%(hour) \\
\hline
$CO_g$&Operational cost of generator $g$ \\%($\$/MWh$)\\
\hline
$P^{\sf max}_g$ & Maximum power generated by unit $g$ \\%($MW$) \\
\hline
$P^{P}_{b,d}$ & Forecast demand at bus $b$ in period $d$ \\%($MW$) \\
\hline
$wp^{P}_{b,d}$ & Forecast wind power at bus $b$ at time $d$ \\%($MW$)\\

\hline
$pv^{P}_{b,d}$ & Forecast PV power at bus $b$ at time $d$ \\%($MW$)\\
\hline
$G_c, \beta_c, R_c$ & Conductance, susceptance, and resistance of line $c$ \\%($\Omega$) \\% (P.U., P.U., $\Omega$) \\
\hline
% $R_c$ & Resistance of line c \\% ($\Omega$)\\
%  \hline
% $\beta_c$ & Susceptance of line $c$ (P.U.)\\
% \hline
$EV^{P}_{b,d}$ & Forecast charging load at bus $b$ at time $d$ \\% ($MW$)\\
\hline
$pf^{max}_c$& Maximum power transferred through line $c$ \\ % ($MW$) \\
\hline
$T^{max}_c$ & Maximum endurable temperature for line $c$ \\%($K$)\\
\hline
$\lambda_c$ & Heat capacity of line $c$ \\%($MJ/mK$)\\
\hline
$T^{E}_{c,d}$ & Temperature of environment around line $c$ at period $d$ \\% ($K$)\\

\hline
$Gs_{c,d}$ & Solar radiation \\%($MW/m^{2}$)\\
\hline
$K^{s}_{c,d}$ & Solar radiation heat gain coefficient of line $c$ at time $d$ \\%($m$)\\
\hline
$D_c$ & Diameter of conductor \\%($m$)\\
\hline
$v_{c,d}$ & Wind velocity at time $d$\\%($m/s$)\\
\hline
$H_c$ & Height of lines above sea \\
\hline
$\rho_{c}$ & Air density around conductor $c$ \\%($kg/m^3$)\\
\hline
$\nu_c$ & Dynamic viscosity of
air around conductor $c$ \\%($kg/m.s$)\\
\hline
$\gamma^{con}_c$& Coefficient of thermal conductivity of air  \\%($W/m.^{o}C$)\\
\hline
$A$& Coefficient related to angle of direction of wind flow and\\& the axis of the conductor\\
\hline
$K^{r}_{c,d}$ &Coefficient of radiation heat loss for line $c$ in period $d$\\% ($MW/mK^{4}$)\\
\hline
$T_{c}^{\sf ref}$ &Reference temperature of line $c$ \\%($K$)\\
\hline
$R_{c}^{\sf ref}$ &Resistance of line $c$ at reference temperature \\%($\Omega$)\\
\hline
$\hbar_c$ &Coefficient of thermal resistivity related to line $c$ \\%($\Omega/K$)\\
\hline
$\epsilon$ &Weather emissivity\\
\hline
\multicolumn{2}{|c|}{\textbf{Variables}}\\

\hline	
$u_c$ & Binary variable, $``1$'' if a new line $c$ is added\\
\hline
$u_g$ &Binary variable, $``1$'' if a new generator $g$ is added\\
\hline
$P_{g,d}$ & Power generated by generator $g$ in period $d$ \\%(MW)\\
\hline
$\alpha_{b,d}$ &Voltage angle of bus $b$ in period $d$\\
\hline
$pf_{c,d}$ &Transferred power through line $c$ in period $d$ \\ %(MW)\\
\hline
$|I_{c,d}|$ &Current magnitude of line $c$ in period $d$ \\ %%(A)\\
\hline
$T_{c,d}$ &Temperature of line $c$ in period $d$ \\%(K)\\

\hline

\end{tabular} \label{notation}
\end{table}

The maximum flow capacity of a line $c$ is $pf^{max}_c$ and the highest temperature it can endure is $T^{max}_c$. % should be satisfied. 
The DTLR parameters include: heat capacity $\lambda_c$, environmental temperature $T^{E}_{c,d}$, solar radiation $Gs_{c,d}$, solar radiation heat gain coefficient $K^{s}_{c,d}$, conductor diameter $D_c$, wind speed $v_{c,d}$, elevation of the line above sea level $H_c$, air density $\rho_c$,
dynamic viscosity of air $\nu_c$,
coefficient of thermal conductivity $k^{c}_c$, coefficient related to the angle between the wind flow direction and the conductor’s axis $A$, coefficient of radiation heat loss
$K^{r}_{c,d}$, reference temperature $T_{c}^{\sf ref}$, resistance at reference temperature $R_{c}^{\sf ref}$, coefficient of thermal resistivity
$\hbar_c$, and  weather emissivity $\epsilon$. $\alpha_{b,d}$ indicates the voltage angle, and the flow of line $c$ in period $d$ is represented by $pf_{c,d}$. Also, let $T_{c,d}$ denote the line temperature.  The key notations are provided in Table \ref{notation}.
%The problem is examined in two scenarios: one without DTLR and one with DTLR, with uncertainty considered in both cases, and applied to two case studies.
%\vspace{0.2in}

%\vspace{0in}
\section{Problem Formulation}
\label{sec:formu}
%In this section, we first present the deterministic formulation of the IGTEP problem. Next, we introduce the hybrid stochastic/robust optimization method to transform the deterministic model into a stochastic/robust model. Finally, we present the IGTEP problem under uncertainty. %The IGTEP problem is defined in this section. The objective function including investment and operation costs should be minimized, while technical and operational constraints will be met.

\subsection{Deterministic IGTEP Problem without DTLR}
\label{deterministic}

In the IGTEP problem, the network operator seeks to minimize the total investment and operational costs, which are captured by the following objective function
\cite{meza2009multiobjective},\cite{hedman2010co}:
% \begin{align}
% &\underset{u_g,u_c,P_{g,d}}{\text{min}} \sum_{c\in \Upsilon_{L}^{\sf  N}}   u_c NC_c+\sum_{g\in \Upsilon{g,n}} u_g NC_g \label{objective} \\ & +\sum_{d \in P_d} \sum_{g \in \Upsilon_{G}} P_{g,d} CO_g \Delta_d \nonumber
% % \label{objective}
% \end{align}

\begin{align}
\label{objective}
    & \min_{u_g \!,u_c \!,P_{\! g,d}} \! \! \sum_{c \in \Upsilon_{\! L}^{\sf N}} \! \! \! \! u_c \text{NC}_c \! + \! \! \sum_{g \in \! \Upsilon_{\! G}^{\sf  N}} \! \! \! \! u_g \text{NC}_{\!g} \!  + \! \! \sum_{d \in \! P_{\!d}} \sum_{g \in \! \Upsilon_{\! G}} \! \! P_{\!g,d} \text{CO}_{\!g} \Delta_d 
\end{align}

\vspace{-0.0cm}
The operator's objective function comprises three components. The first two terms in (\ref{objective}) represent the installation costs of new transmission lines and new generators, respectively. The third term accounts for the operational costs of the generators over the entire planning horizon. The operator's planning and operations are subject to the following constraints:
% All of these terms should be minimized to achieve the lowest investment and operational costs.  Constraints are represented as follows:
\begin{align}
    &u_c=1, \quad c \in \{\Upsilon_{L} \setminus \Upsilon_{L}^{\sf  N}\} \label{existing line binary} \\
    & u_g=1, \quad g \in \{\Upsilon_{G} \setminus \Upsilon_{G}^{\sf N}\}
    \label{existing unit binary} \\
    & u_c \in \{0,1\}, \quad c \in \Upsilon_{L}^{\sf  N}
    \label{new line binary} \\
    &u_g \in \{0,1\}, \quad g \in \Upsilon_{G}^{\sf N}
 \label{new unit binary}\\
    & 0 \leq P_{g,d} \leq u_g P^{max}_g, \quad \forall g, ~\forall d
    \label{unit capacity} \\
    & \sum_{g \in \mho^{g}_b} P_{g,d} - \sum_{c:s(c)=b} pf_{c,d} + \sum_{c:r(c)=b} pf_{c,d} ~ = \nonumber  \\
    &  \left(P^{p}_{b,d} + EV^{p}_{b,d} - wp^{p}_{b,d} -pv^{p}_{b,d} \right), \quad \forall b,~ \forall d \label{balance equation}  \\
    & pf_{c,d} = u_c \beta_c (\alpha_{s(c),d} - \alpha_{r(c),d}), \quad \forall c,~ \forall d \label{DC load flow} \\
    & pf_{c,d} \leq pf^{max}_c, \quad \forall c, ~\forall d \label{max flow} \\
    & pf_{c,d} \geq -pf^{max}_c, \quad \forall c, ~\forall d \label{min flow}
\end{align}

% \begin{equation}
% u_c=1, \quad c \in \{\Upsilon_{L} - \Upsilon_{L}^{\sf  N}\}
% \label{existing line binary}
% \end{equation}

% \begin{equation}
% u_g=1, \quad g \in \{\Upsilon_{G} - \Upsilon_{G}^{\sf N}\}
% \label{existing unit binary}
% \end{equation}

% \begin{equation}
% u_c \in \{0,1\}, \quad c \in \Upsilon_{L}^{\sf  N}
% \label{new line binary}
% \end{equation}

% \begin{equation}
% u_g \in \{0,1\}, \quad g \in \Upsilon_{G}^{\sf N}
% \label{new unit binary}
% \end{equation}

% \begin{equation}
% 0 \leq P_{g,d} \leq u_g P^{max}_g, \quad \forall g, \forall d
% \label{unit capacity}
% \end{equation}

% \begin{equation}
% \label{balance equation}
% \sum_{g \in \mho^{g}_b} P_{g,d} - \sum_{c:s(c)=b} pf_{c,d} + \sum_{c:r(c)=b} pf_{c,d} =\\ \nonumber\sum_{l \in \mho^{l}_b} \left(P^{p}_{l,d} + EV^{p}_{l,d} - wp^{p}_{l,d} - pv^{p}_{l,d}\right), 
% \forall b, \forall d
% \end{equation}

% \begin{equation}
% pf_{c,d} = u_c \beta_c (\alpha_{s(c),d} - \alpha_{r(c),d}), \quad \forall c, \forall d
% \label{DC load flow}
% \end{equation}

% \begin{equation}
% pf_{c,d} \leq pf^{max}_c, \quad \forall c, \forall d
% \label{max flow}
% \end{equation}

% \begin{equation}
% pf_{c,d} \geq -pf^{max}_c, \quad \forall c, \forall d
% \label{min flow}
% \end{equation}

Constraints (\ref{existing line binary}) specify that the binary variable $u_c$ for each existing line is equal to 1. Constraints (\ref{existing unit binary}) state that the binary variable for each % pertaining to the 
existing generator is set to 1. Constraints (\ref{new line binary}) and (\ref{new unit binary}) are related to the binary indicators for installing new lines and generators. If the operator decides to install line $c$ in the set of candidate new lines $\Upsilon_{L}^{\sf  N}$, then $u_c$ takes the value of ``1''. Similarly, if the operator installs a new generator $g$ in the set of candidate new generators $\Upsilon_{G}^{\sf  N}$, then $u_g$ equals ``1''. 
Constraints (\ref{unit capacity}) impose the power output limit of each generator.
%, the power limits of the units are specified. 
The energy balance constraints are defined in (\ref{balance equation}), and the DC load flow equations are provided in (\ref{DC load flow}). Constraints (\ref{max flow}) and (\ref{min flow}) describe the line flow limits. %Consequently, constraints (\ref{existing line binary})-(\ref{min flow}) are considered as deterministic constraints without DTLR.

Overall, constraints (\ref{objective})-(\ref{min flow}) define the deterministic IGTEP problem without considering DTLR. However, this deterministic model does not account for various system uncertainties. %, such as renewable energy generation, load fluctuations, and weather conditions.
Thus, it may lead to suboptimal solutions, significantly impacting system performance during actual operations. To address these uncertainties, we will present the hybrid stochastic robust optimization method. Then, we will introduce the IGTEP problem under uncertainty, both with and without DTLR.

%\vspace{0in}
\subsection{Hybrid Stochastic/Robust Problem Definition}
\label{methodology}

This section presents the hybrid SP/RO method for dealing with various system uncertainties \cite{lin2004new},\cite{janak2007new}. %provides a detailed proof of the SP/RO method. 
We begin by considering the following generic deterministic formulation. %The problem addressed is an MILP problem that incorporates uncertainty \cite{lin2004new},\cite{janak2007new}.
\begin{align}
&\underset{n,m}{\text{min}}~~~ e^{\sf T} n+w^{\sf T} m
\label{obj} \\
& Fn+Pm\leq j \label{original}
\\
&n_{min} \leq n \leq n_{max} \label{range}\\
&m_v=\{0,1\},~~~\forall v \label{range binary}
\end{align}

%The original deterministic problem is represented in (\ref{obj})-(\ref{range binary}). 
In this generic MILP problem, $n$ and $m$ are variables. The parameters include a vector $j$ and matrices $F$ and $P$ with appropriate sizes. % Let $F$ and $P$ be matrices that serves as parameters, and let $j$ be a vector parameter, all of which are parameters subject to uncertainty. 
%Since the parameters can be uncertain. 
With certain assumptions on the probability distribution of the uncertainties, we will show that the deterministic formulation (\ref{obj})-(\ref{range binary}) can be reformulated to deal with uncertain parameters. % into a counterpart problem capturing all the uncertain parameters. 
In particular, we can replace constraint (\ref{original}) with the following constraints \cite{lin2004new,janak2007new}: 
\begin{align}
&\sum_{t} f_{s,t} n_t+ \phi \omega \sqrt{\sum_{t} f_{s,t}^{\sf 2} n_t^{\sf 2}+\sum_{v} p_{s,v}^{\sf 2} m_v+j_s^{\sf 2}}\label{modified} \\ & +\sum_{v} p_{s,v} m_v\leq j_s+\mu \max{\Big[1,|j_s|\Big]},~~~\forall s \nonumber
\end{align}
where $s$, $t$, and $v$ are indices related to the uncertain parameters $F, P,$ an $j$. Additionally, $f_{s,t}, ~p_{s,v},$ and $j_s$ are the forecast values of the uncertain parameters. 
Also, $\phi$, and $\mu$ are defined as level of the uncertainty and infeasibility tolerance, respectively in (\ref{modified}). The expression of $\omega$ will be provided later.
%Also, $\phi$ is multiplied by standard deviation of parameters on the left-hand side of (\ref{modified}).
% $F$ and $P$ are represented by sets of possible values denoted by $T_s$ and $V_s$. 
%To prove the transformation (\ref{modified}), we need
We have the following definitions that indicate the relationship between the predicted values of the parameters and their true values. 
\begin{align}
&f_{s,t}^{\sf true}=(1+\phi \psi_{s,t}) f_{s,t}
\label{true valuef}\\
&p_{s,v}^{\sf true}=(1+\phi \psi_{s,v}) p_{s,v}
\label{true valuep}\\
&j_{s}^{\sf true}=(1+\phi \psi_{s}) j_{s}, \label{true valuej}
\end{align}
%(\ref{true valuef})-(\ref{true valuej}) indicate relationship between nominal values of parameters and true values. 
where $f_{s,t}^{\sf true}, p_{s,v}^{\sf true},$ and $j_{s}^{\sf true}$ are the actual realization of the uncertainties. We use $\phi$ and $\psi$ to capture the deviation of the uncertain parameters from their forecast values. We assume the uncertain parameters follow normal distributions. Thus, $\psi$ follows normal distributions. We will justify this assumption in Section \ref{uncertainty}. 
%Variable $\psi$ represents a Normal probability function.
%To validate the inclusion of 
For constraints (\ref{modified}) to hold, the following two conditions must be satisfied \cite{lin2004new,janak2007new}:
(i) The deterministic problem, which uses the forecast values for the uncertainties, is feasible; and ii) %The problem must be feasible when using the forecast values for the uncertain parameters.
%
% (I): The problem must be feasible when using the forecast values for the uncertain parameters.
The probability of constraint violation is small. We can enforce:
\begin{align}
&\Pr \left \{\sum_{t} f_{s,t}^{\sf true} n_t+\sum_{v} p_{s,v}^{\sf true} m_v > j_s^{\sf true} +\mu \max{[1,|j_s|]} \right \}  \leq \Re \nonumber
\end{align}
where $\omega=F_n^{-1}(1-\Re)$ and $\Re$ is reliability level. 
%\end{itemize}
We can also express the relationship between $\omega$ and $\Re$ as
\begin{align}
\label{reliability}
\Re=1-\int\limits_{-\infty}^{\omega} \frac{1}{\sqrt {2\pi}} e^{\sf \frac {-x^{\sf 2}} {2}}\,dx. 
\end{align}
%In fact, integral function in (\ref{reliability}) is the definition of $F_n(\omega)$.

% Proof (II):
% \begin{align}
% & \Pr \left\{ 
% \begin{aligned}
% & \sum_{t} f_{s,t}^{\sf true} n_t + \sum_{v} p_{s,v}^{\sf true} m_v \\
% & > j_s^{\sf true} + \mu \max \left[1, |j_s|\right]
% \end{aligned}
% \right\} \nonumber \\
% & = \Pr \left\{ 
% \begin{aligned}
% & \sum_{t} f_{s,t} n_t + \phi \sum_{t \in T_S} \psi_{s,t} |f_{s,t}| n_t \\
% & + \sum_{v} p_{s,v} m_v + \phi \sum_{v \in V_s} \psi_{s,v} |p_{s,v}| m_v \\
% & > j_s + \phi \psi_s |j_s| + \mu \max \left[1, |j_s|\right]
% \end{aligned}
% \right\}\leq \nonumber \\
% &  \Pr \left\{ 
% \underbrace{\frac{\sum_{t} \psi_{s,t} |f_{s,t}| n_t + \sum_{v} \psi_{s,v} |p_{s,v}| m_v - \psi_s |j_s|}{\sqrt{\sum_{t} f_{s,t}^{\sf 2} n_t^{\sf 2} + \sum_{v} p_{s,v}^{\sf 2} m_v + j_s^{\sf 2}}}}_\text{$\lambda$} > \omega
% \right\}  \nonumber \\
% & = 1 - \Pr \left\{\lambda \leq \omega \right \}  \nonumber \\
% & = 1 - F_n(\omega) \nonumber \\
% & = 1 - (1 - \Re) = \Re \nonumber
% \end{align}

% where $\lambda$ is a random
% variable with standardized Normal distribution function. Also, relationship between $\omega$ and $\Re$ is represented below:
% \begin{align}
% \label{reliability}
% \Re=1-\int\limits_{-\infty}^{\omega} \frac{1}{\sqrt {2\pi}} e^{\sf \frac {-x^{\sf 2}} {2}}\,dx 
% \end{align}
% In fact, integral function in (\ref{reliability}) is the definition of $F_n(\omega)$.

\subsection{Probability Distribution Justifications}
\label{uncertainty}
%To apply the approach that combines RO and (SP), uncertain parameters must be defined using probability distribution functions. 
In our problem, some uncertainties may not follow the normal distributions. While the base loads are typically represented by Normal Probability Distribution Functions (NPDF), the wind power generation, wind speed, and solar irradiance are often characterized by other distributions such as Weibull Probability Distribution Functions (WPDFs) \cite{khamees2022mixture}. Approximating WPDFs with NPDFs is an effective method for addressing this problem \cite{baharvandi2018bundled}. Similarly, the output power of photovoltaic systems (PVs) is usually modeled by Beta Probability Distribution Functions (BPDFs), which can be approximated by NPDFs with minimal errors \cite{baharvandi2019linearized}. %Therefore, we can use approximation techniques to approximate 
%the uncertainty implementation for all parameters was conducted using the proposed method outlined in the previous section. 
The charging load profile of EVs depends on three main factors: (i)	When EVs are connected to the network; (ii) How much power they need; and (iii) The initial state of charge before being connected.
When EV data are available, the Probability Density Function (PDF) related to EV charging load can be defined. %Since all EVs are similar in terms of their charging power requirements and connection probabilities, 
We use a binomial distribution function to model these connection probabilities of EVs. This distribution is appropriate for events with two possible states—connected to or disconnected from the network—mirroring the binary nature of EV connectivity. The binomial distribution is defined as follows:
\begin{align}
\text{prob}(x) = \frac{n!}{x!(n-x)!} \rho^x q^{n-x}, \quad \text{for } x = 0, 1, 2, \ldots, n
\label{binomial}
\end{align}
In this distribution, $x$ represents the number of EVs connected to the network. The parameter $\rho$ denotes the probability of an EV being connected, while $q$ is the probability of an EV not being connected. The parameter $n$ represents the total number of EVs.
When the number of EVs is large, the binomial distribution can be approximated by an NPDF \cite{binomial}. Specifically, as the number of charging EVs increases, the corresponding binomial distribution approaches a normal distribution.  Thus, in this work, all uncertain parameters are assumed to follow normal distributions, allowing for the effective application of the proposed method to manage uncertainty.

\subsection{IGTEP Formulation under Uncertainty}

Building on the method outlined in Section \ref{methodology}, we now introduce the IGTEP formulations under uncertainty, both with and without the consideration of DTLR. % We first provide the IGTEP model     The uncertain problem will be defined in two subsections: one with DTLR and one without.

\subsubsection{Uncertainty-aware IGTEP without DTLR}
\label{without DTLR}
Instead of considering separately the uncertainties in the base load, EV charging load, and renewable energy generation, we group these uncertainties into a single uncertain parameter expressing the net load on the right-hand side of constraints (\ref{balance equation}). Subsequently, we can employ the results presented in Section  \ref{methodology}.
% To use the results presented in %To capture uncertainty of demand, EV charging load, PVs, and wind turbines, constraint (\ref{balance equation}) should be modified based on section
% Section \ref{methodology}, we grouped together in a single term on the right-hand side of constraints (\ref{balance equation}).
In particular, constraints (\ref{balance equation}) can be reformulated to account for uncertainties as follows: %in the base load, EV charging load, and renewable energy generation. %The modified constraints are as follows:
%. First, we relax this constraint. After incorporating uncertainty, the following constraint should be applied:
%\vspace{0cm}
\begin{align}
&P^{net}_{b,d}= P^{p}_{b,d}+EV^{p}_{b,d}-wp^{p}_{b,d}-pv^{p}_{b,d}\ ~~~\forall b, ~\forall d \label{Pnet}
\\
&\sum_{g\in \mho^{g}_b} P_{g,d}-\sum_{c:s(c)=b} pf_{c,d}+\sum_{c:r(c)=b} pf_{c,d} \geq \label{modified BE} \\ &  \Big( P^{net}_{b,d} \Big) -\mu \max \Big[1,\Big| \Big(P^{net}_{b,d}\Big) \Big| \Big] +\phi \omega \Big[ \Big(P^{net}_{b,d}\Big)\Big],~~~\forall b, ~\forall d \nonumber % \in \mho_b,\forall d \nonumber
\end{align}
where $P^{net}_{b,d}$ in (\ref{Pnet}) expresses the right hand side of (\ref{balance equation}), which is the uncertain net demand.  %is defined as net demand facing uncertainty. 
Note that $\mu$ and $\phi$ are defined as infeasibility tolerance and uncertainty level, respectively in (\ref{modified BE}). Furthermore, relationship between $\omega$ and reliability level ($\Re$) is described in (\ref{reliability}).

\subsubsection{Uncertainty-aware IGTEP with DTLR}
\label{with DTLR}
The capacity of overhead transmission lines (OTLs) plays a crucial role in the planning and operations of power systems. DTLR is used to determine this capacity. The assessment of DTLR is conducted through three different approaches:
\begin{itemize}
    \item Prediction of DTLR by forecasting the demand of network and weather conditions \cite{kim2006prediction}, \cite{siwy2006risk}.

    \item Indirect measurements \cite{mai2011dynamic}.

    \item  Real-time assessment of DTLR from the data acquired by meteorologists \cite{foss1983dynamic}.
\end{itemize}

In practice, the heat balance equation (HBE) is used to calculate the capacity of the lines, which is expressed as %follows: %and it is formulated as follows \cite{???}:
\begin{align}
Q_{gain}=Q_{release},
\label{energy balance}
\end{align}
%This equation includes four terms that will be explained in detail.
where $Q_{gain}$ includes ohmic losses and solar heat gains, while $Q_{release}$ includes convection heat losses and radiation heat losses. We will elaborate on these terms in the following. First, the ohmic losses are given below: % presented in (\ref{ohmic}):
\begin{align}
Q^{o}_{c,d}(T_{c,d})=R(T_{c,d})|I_{c,d}|^{2},~~~\forall c,~\forall d
\label{ohmic}
\end{align}
where $R(T_{c,d})$ and $I_{c,d}$ are the resistance and the current of line $c$, respectively. %of line $c$ Also, $I_{c,d}$ represents the current of line $c$. 
The solar heat gained by line $c$ in period $d$ depends on three parameters, as shown below:
\begin{align}
Q^{s}_{c,d}=u_c K^{s}_{c,d} D_c Gs_{c,d},~~~\forall c,~\forall d
\label{solar}
\end{align}
The line temperature increases due to the effects described in equations (\ref{ohmic}) and (\ref{solar}). %The third term
The convection heat losses in the HBE depend on the line temperature and weather conditions, such as surrounding temperature, altitude, and speed of wind. %This term represents convection heat losses. 
We have:
\begin{align}
&Q^{con}_{c,d}(T_{c,d})=u_c f^{NW}_{c,d}(v_{c,d},T^{E}_{c,d},D_c,H_c)\label{convection}\\&(T_{c,d}-T^{E}_{c,d})^{1.25}, \nonumber~~~\forall c,~\forall d \nonumber
\end{align}
where $f^{NW}_{c,d}$ is a function in the convection formula for no wind conditions. Finally, the radiation heat losses are: % can be given as % represented in (\ref{radiation}):
\begin{align}
Q^{rad}_{c,d}(T_{c,d})=u_c \epsilon K^{r}_{c,d}((T_{c,d})^{4}-(T^{E}_{c,d})^{4}),~~~\forall c,~\forall d
\label{radiation}
\end{align}
Convection and radiation heat losses help reduce the temperature of the transmission lines. As a result, system operators can more accurately determine the maximum power flow in the lines using the HBE method. Consequently, the following constraints should be incorporated into the IGTEP problem:
\begin{align}
&Q^{o}_{c,d}(T_{c,d})+ Q^{s}_{c,d}=Q^{con}_{c,d}(T_{c,d})+Q^{rad}_{c,d}(T_{c,d}),~\forall c,\forall d \label{HBE}\\
&T_{c,d}\leq u_c T^{max}_c, ~\forall c
\label{Tmax}\\
&pf_{c,d}=u_c(G_c-G_c \cos(\alpha_{s(c),d}-\alpha_{r(c),d}), ~\forall c,~\forall d \label{AC load flow}\\ &\nonumber +\beta_c \sin(\alpha_{s(c),d}-\alpha_{r(c),d})),~~~\forall c,~\forall d \nonumber \\
&R(T_{c,d})=R_{c}^{\sf ref}(1+\hbar(T_{c,d}-T_{c}^{\sf ref})),~~~\forall c,~\forall d
\label{R(Tcd)} \\
&V_b I_{c,d}=|pf_{c,d}|,~~~\forall c,~\forall d
\label{active power formula}\\
&Q^{o}_{c,d}= R(T^{max}_c)I_{c,d}^{2},~~~\forall c,~\forall d \label{ohmic definition}
\end{align}

%To deal with a simplified problem,
Note that for simplicity, we assume the amplitude of bus voltage is 1 P.U. and we also define $R(T^{max}_c)$. The HBE is detailed in (\ref{HBE}). Constraints (\ref{Tmax}) specify the maximum allowable line temperature. Equation (\ref{AC load flow}) states the power flow of the line in the AC load flow model. Utilizing AC load flow in the context of DTLR is crucial because ohmic losses, which depend on the line resistance, must be accounted for. Thus, the DC load flow would be insufficient. Consequently, constraints (\ref{DC load flow})-({\ref{min flow}}) do not apply to the DTLR model and should be replaced by constraints (\ref{HBE})-(\ref{AC load flow}). The relationship between the resistance of transmission lines and their temperature is given by equation (\ref{R(Tcd)}). According to equation (\ref{active power formula}), the line current and active power flow are assumed to be equivalent. By substituting the active power flow for the current in the HBE, the relationship between active power flow and line temperature can be established. Utilizing $R(T_c^{\sf max})$ in (\ref{ohmic}) simplifies the problem outlined in (\ref{ohmic definition}).

To evaluate wind speed in convection heat, equation (\ref{convection}) should be adjusted as follows:
\begin{align}
Q^{con1}_{c,d}(T_{c,d})=u_c A (1.01+1.35(R_{c,d}^{Re})^{0.52})\gamma^{con}_c (T_{c,d}-T^{E}_{c,d}) \nonumber \\ 
~~ \forall c,~\forall d    \nonumber  \\
%\quad \quad \quad \quad \quad \quad \quad (T_{c,d}-T^{E}_{c,d}),~~ \forall c,~\forall d \nonumber \\
Q^{con2}_{c,d}(T_{c,d})=u_c A (0.754)(R_{c,d}^{Re})^{0.5}\gamma^{con}_c (T_{c,d}-T^{E}_{c,d}),  \forall c,\forall d    \nonumber \\ 
%(T_{c,d}-T^{E}_{c,d}),~~  \forall c,~\forall d \nonumber \\
R_{c,d}^{Re}=\frac{D_c v_{c,d} \rho_c}{\nu_c},~~  \forall c,~\forall d  \nonumber 
\end{align}
where $R_{c,d}^{Re}$ represents the Reynolds number, which is dimensionless. Note that equation (\ref{convection}) describes the natural convection, which occurs in the absence of wind. In contrast, these equations %(\ref{convection1}) and (\ref{convection2})
pertain to the forced convection, which is influenced by wind speed. The highest value among these should be selected for calculation in the HBE. To incorporate wind speed uncertainty, simplifications need to be applied to these equations. % (\ref{convection1}) and (\ref{convection2}). 
After making these adjustments, the RO/SP method can be used as follows:
\begin{align}
&k'_{c,d}= A (1.01+1.35(R_{c,d}^{Re})^{0.52})\gamma^{con}_c,~~~\forall c,~\forall d \label{kp} \\
&k''_{c,d}= A (0.754)(R_{c,d}^{Re})^{0.5}\gamma^{con}_c,~~~\forall c,~\forall d \label{kpp} \\
&T'_{c,d}=T_{c,d}-T^{E}_{c,d},~~~\forall c,~\forall d \label{Tp} \\
&Q^{con1}_{c,d}(T'_{c,d})+\phi \omega u_c k'_{c,d}T'_{c,d}-u_c k'_{c,d}T'_{c,d}\leq \mu,\forall c,\forall d \label{uncertain convection1} \\
&Q^{con2}_{c,d}(T'_{c,d})+\phi \omega u_c k''_{c,d}T'_{c,d}-u_c k''_{c,d}T'_{c,d}\leq \mu,\forall c,\forall d \label{uncertain convection2}
\end{align}
where (\ref{kp})-(\ref{Tp}) are the simplifications. Constraints (\ref{uncertain convection1}) and (\ref{uncertain convection2}) state the convection heat loss equations %(\ref{convection1}) and (\ref{convection2}) 
after relaxing and implementing uncertainty of $k'_{c,d}$ and $k''_{c,d}$. Thus, the HBE will be modified as follows:
\begin{align}
&Q^{o}_{c,d}(T_{c,d})+Q^{s}_{c,d}=\max\{Q^{con1}_{c,d}(T'_{c,d}), Q^{con2}_{c,d}(T'_{c,d})\} \label{max convection}\\ & +Q^{rad}_{c,d}(T_{c,d}),~~~\forall c,~\forall d \nonumber
\end{align}
The maximum term in (\ref{max convection}) introduces non-linearity into the model. To address this non-linearity, we introduce a binary variable, $y_{c,d}$
, along with two additional constraints for each conductor and each period. These constraints are as follows:
\begin{align}
&\max\{Q^{con1}_{c,d}(T'_{c,d}), Q^{con2}_{c,d}(T'_{c,d})\}=y_{c,d}Q^{con1}_{c,d}(T'_{c,d}) \label{max term}\\ & \quad \quad \quad \quad \quad \quad +(1-y_{c,d})Q^{con2}_{c,d}(T'_{c,d}),~~\forall c,~\forall d \nonumber \\
&Q^{con1}_{c,d}(T'_{c,d})\geq Q^{con2}_{c,d}(T'_{c,d})y_{c,d},~~\forall c,~\forall d \label{max1} \\
&Q^{con1}_{c,d}(T'_{c,d})(1-y_{c,d})\leq Q^{con2}_{c,d}(T'_{c,d}),~~\forall c,~\forall d \label{max2}
\end{align}
To linearize equations (\ref{max term})--(\ref{max2}), additional constraints and variables are required, which can complicate the problem. To simplify the model and avoid introducing these extra variables and constraints through the Big-M method, the following approach can be utilized:
\begin{align}
&k'_{c,d}\geq k''_{c,d}y_{c,d},~~~\forall c,~\forall d \label{kpgkpp} \\
&k'_{c,d}(1-y_{c,d})\leq k''_{c,d},~~~\forall c,~\forall d \label{kplkpp}\\
&Q^{con1}_{c,d}(T'_{c,d})\leq M y_{c,d},~~~\forall c,~\forall d \label{con1ly}\\
&Q^{con2}_{c,d}(T'_{c,d})\leq M (1-y_{c,d}),~~~\forall c,~\forall d \label{con2ly} \\
&Q^{o}_{c,d}(T_{c,d})+Q^{s}_{c,d}=Q^{con1}_{c,d}(T'_{c,d})+Q^{con2}_{c,d}(T'_{c,d}) \label{fHBE}\\ &+Q^{rad}_{c,d}(T_{c,d}),~~~\forall c,~\forall d\nonumber
\end{align}
According to (\ref{kpgkpp}), when $y_{c,d}=1$, so $k'_{c,d}$ is greater than $k''_{c,d}$ which means $Q^{con1}_{c,d}(T'_{c,d})$ is greater than $Q^{con2}_{c,d}(T'_{c,d})$ based on the convection heat loss equations. % (\ref{convection1}) and (\ref{convection2}), 
and (\ref{kplkpp}) states that $k''_{c,d}\geq 0$. Constraints (\ref{con1ly}) show that $Q^{con1}_{c,d}(T'_{c,d})$ should be less than a sufficiently big number M for $y_{c,d}=1$, and constraints (\ref{con2ly}) indicate $Q^{con2}_{c,d}(T'_{c,d})\leq 0$, while convection heat should be a non-negative variable. Thus, it will be zero. Based on the HBE equation defined in (\ref{fHBE}), only $Q^{con1}_{c,d}(T'_{c,d})$ will appear which is the maximum. Suppose $y_{c,d}=0$, in (\ref{kpgkpp}), $k'_{c,d}\geq 0$, and in (\ref{kplkpp}), $k''_{c,d}$ is more than $k'_{c,d}$ resulting in $Q^{con2}_{c,d}(T'_{c,d})\geq Q^{con1}_{c,d}(T'_{c,d})$. Based on (\ref{con1ly}), $Q^{con1}_{c,d}(T'_{c,d})$ is less than or equal to zero, while it should be non-negative which means it will be zero. Equation  (\ref{con2ly}) states that $Q^{con2}_{c,d}(T'_{c,d})$  will be less than a big number. Consequently, in the HBE equation (\ref{fHBE}), $Q^{con1}_{c,d}(T'_{c,d})$ will not appear, and only the highest convection heat will affect HBE. As a result, instead of (\ref{max term})-(\ref{max2}) and the constraints and variables related to the big-M method that should be applied to linearize them, (\ref{kpgkpp})-(\ref{fHBE}) will be used. 

In equation (\ref{fHBE}), $Q_{c,d}^{s}$ confronts uncertainty since the solar irradiance is uncertain. Eventually, the HBE equation after considering the uncertainty in solar can be given as follows:
\begin{align}
&Q^{o}_{c,d}(T_{c,d})+Q^{s}_{c,d}-\mu \max \{1,|Q^{s}_{c,d}|\}+\phi \omega Q^{s}_{c,d} \label{final uncertain HBE} \\ &\leq Q^{con1}_{c,d}(T'_{c,d})+Q^{con2}_{c,d}(T'_{c,d})+Q^{rad}_{c,d}(T_{c,d}),~~~\forall c,~\forall d \nonumber
\end{align}

It is worth noting that integrating additional uncertainties, such as ambient temperature, into our model is feasible. However, in this study, the impact of ambient temperature fluctuations is minimal because each period is largely representative of a single season. Consequently, the effect of temperature variability on the model's outcomes is negligible. %, and therefore, its uncertainty has been omitted from our analysis.

\section{Linearization Techniques}
\label{Linearization}
We will present the linearization methods to simplify our problem. 
% The term $\max\{a,b\}$ can be rewritten as: $z = a = $
First, (\ref{modified BE}) includes an absolute term that is non-linear. The following formulations can be applied to linearize:
\begin{align}
&\text{max}[1,|\delta|]=\vartheta_1+(\vartheta_2(-\delta)+\vartheta_3(\delta))\vartheta_4
\label{max}\\
&\vartheta_3 \delta \geq 0
\label{xg0}\\
&\vartheta_2\delta\leq 0
\label{xl0}\\
&\vartheta_4(\vartheta_2(-\delta)+\vartheta_3(\delta))\geq \vartheta_4
\label{y4y4}\\
&\vartheta_1(\vartheta_2(-\delta)+\vartheta_3(\delta))\leq \vartheta_1
\label{y1y1}\\
&\vartheta_2+\vartheta_3=1
\label{y2y3}\\
&\vartheta_1+\vartheta_4=1,
\label{y1y4}
\end{align}
where $\vartheta_1$, $\vartheta_2$, $\vartheta_3$, and $\vartheta_4$ are binary variables. If $\vartheta_1$=1, $\vartheta_4$ is zero based on (\ref{y1y4}), and constraint (\ref{y4y4}) will be inactive. In this situation, constraint (\ref{y1y1}) states $|\delta|\leq 1$, so equation (\ref{max}) will be equal to 1. On the other hand, if $\vartheta_1$=0, $\vartheta_4$ is 1 according to (\ref{y1y4}), and the constraint (\ref{y1y1}) is inactive. The constraint (\ref{y4y4}) indicates $|\delta|\geq 1$. In this case, if $\vartheta_3$=1, $\vartheta_2$ will be 0 according to (\ref{y2y3}). As a result, constraint (\ref{xl0}) gets inactive, and constraint (\ref{xg0}) shows $\delta$ is non-negative, so the equation (\ref{max}) is equal to $\delta$. In contrast, if $\vartheta_2$=1, $\vartheta_3$ is zero based on (\ref{y2y3}), and the constraint (\ref{xg0}) is inactive, while $\delta$ is non-positive as shown in the constraint (\ref{xl0}). Thus, the equation (\ref{max}) equals $-\delta$. Non-linearity due to product of variables can be seen in (\ref{max})-(\ref{y1y1}). To linearize constraint (\ref{xg0}), the following constraints (big-M method) can be employed:
\begin{align}
&\vartheta_3 \delta=\theta
\label{main}\\
&\theta \geq -\vartheta_3\times X_1
\label{sgX}\\
&\theta \leq \vartheta_3\times X_1
\label{slX}\\
&\theta \geq \delta-(1-\vartheta_3)\times X_1
\label{sgy3}\\
&\theta \leq \delta+(1-\vartheta_3)\times X_1;
\label{sly3}
\end{align}
 where $X_1$ is a large number in (\ref{sgy3}) and (\ref{sly3}). If $\vartheta_3$=0, based on (\ref{sgX}) and (\ref{slX}), $\theta$=0, and constraints (\ref{sgy3}) and (\ref{sly3}) state $\theta$ can be real value. If $\vartheta_3$=1, constraints \ref{sgy3}) and (\ref{sly3}) state $\theta$=$\delta$, and constraints (\ref{sgX}) and (\ref{slX}) show $\theta$ belongs to real numbers. This linearization should be used for $\vartheta_2(-\delta)$, and multiplication of $\vartheta_4$ and terms in parenthesis as well as other constraints to obtain linear constraints.
Constraints (\ref{DC load flow})-(\ref{min flow}) can be replaced as follows to remove the nonlinearity of constraints (\ref{DC load flow}):
\begin{align}
&-u_c pf^{max}\leq pf_{c,d}\leq u_c pf^{max},~~~\forall c,~\forall d
\label{rangepf}\\
&-(1-u_c)X\leq \frac{pf_{c,d}}{\beta_c}-(\alpha_{s(c),d}-\alpha_{r(c),d}) \label{rangepfX}\\ &\leq (1-u_c)X~~~\forall c, \forall d \nonumber
\end{align}
If $u_c$ equals 0, (\ref{rangepf}) implies that $pf_{c,d}$ will be zero, while for constructed lines, constraint (\ref{rangepfX}) is inactive and states an equality without any binary variable that is the DC load flow equation. 
To linearize trigonometric functions used in AC power flow (\ref{AC load flow}), the following method can be employed:
\begin{align}
&C_i=|\cos(x)-(s_i x+m_i)|
\label{error}\\
&\partial C_i/\partial x=0, \sin(x)=-s_i
\label{derivative}\\
&\cos(x)\approx(1-l)(s_1 x+m_1)+l(s_2 x+m_2)
\label{cosline}\\
&x\geq -0.6 (1-l)
\label{xrange}\\
&x\leq 0.6 l \label{xmin}
\end{align}
To approximate the cosine function over the range of angle differences between voltages, two linear segments are used: one for the interval from 
$-0.6$ to 0 and another for 0 to $0.6$. This range is suitable for angle differences in transmission networks. Equation (\ref{error}) describes the deviation between the linear approximation and the cosine function, while equation (\ref{derivative}) sets the derivative of equation (\ref{error}) to zero to find the maximum value of $C_i$ for many $s_i$ and $m_i$. Also, for some $s$, the value of $C_i$ at $-0.6$ or 0.6 may exceed the value at the extremum point, if one exists. Therefore, for each line used in equation (\ref{error}), we first compare 
$C_i$ with the extremum point value, choosing the maximum among these. 

\begin{figure}[h!]
	\centering
		\includegraphics[width=0.4\textwidth,height=0.10\textheight]{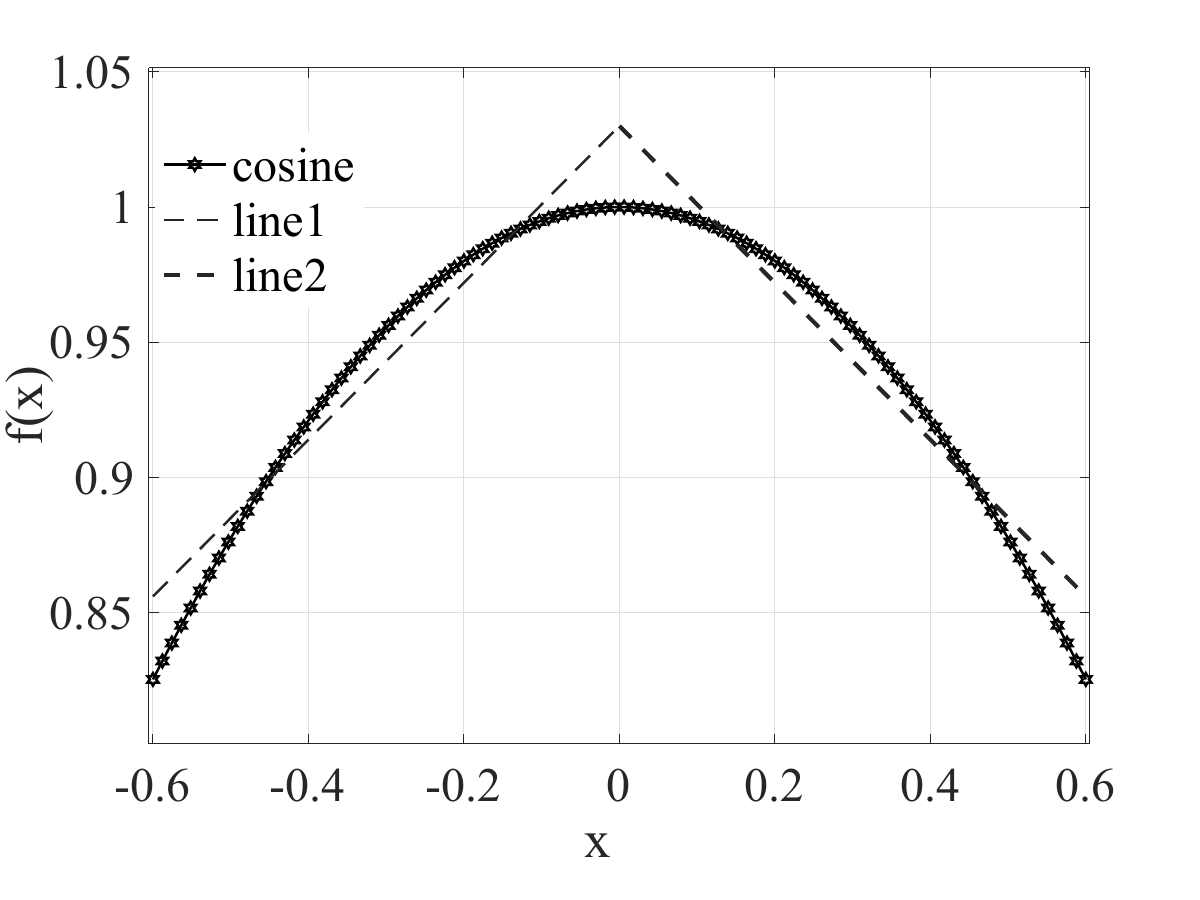}
			\caption{Approximation of cosine function to line}
	\label{fig. 2}
\end{figure}

\begin{figure}[h!]
	\centering
		\includegraphics[width=0.4\textwidth,height=0.10\textheight]{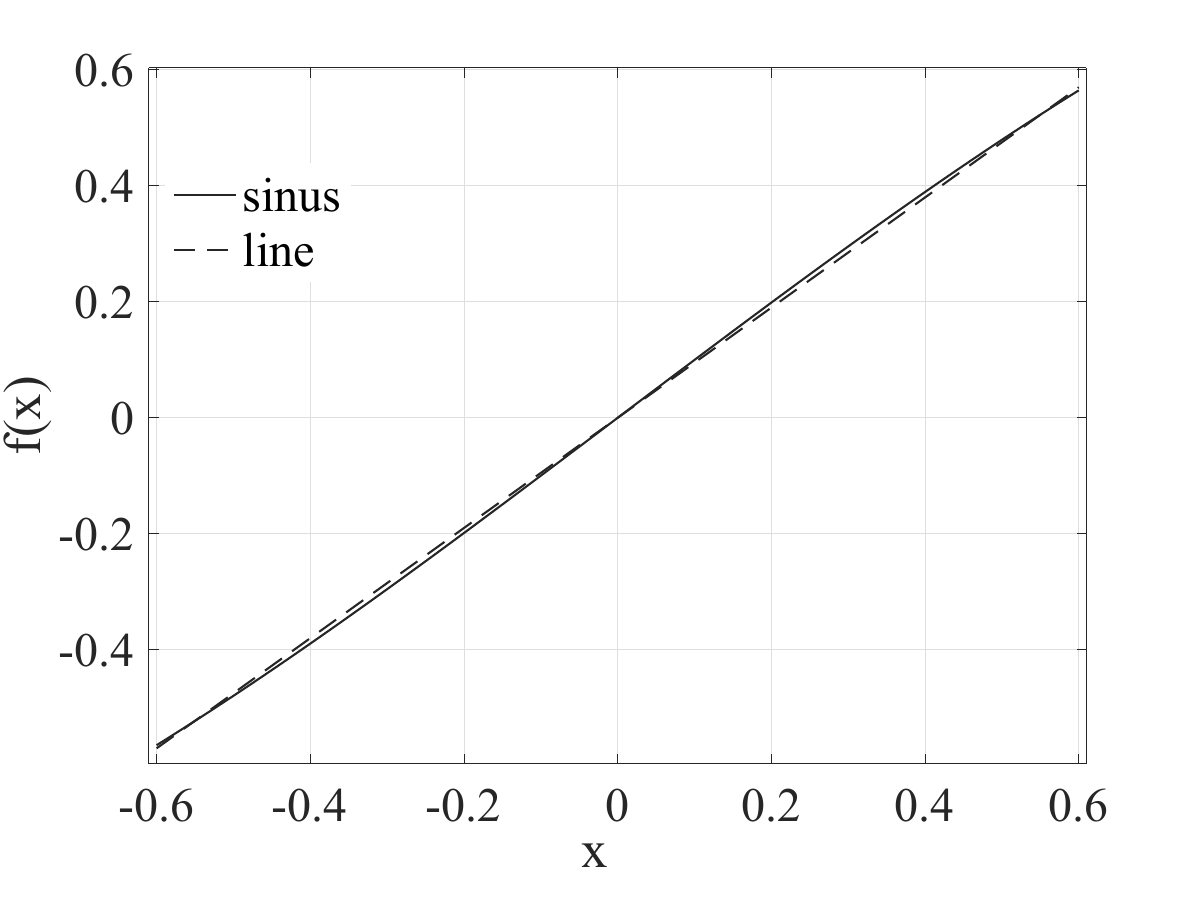}
			\caption{Approximation of sinus function to line}
	\label{fig. 3}
\end{figure}

By applying this process to each line, we determine the maximum deviation between the cosine function and its linear approximations. The smallest of these maximum deviations will provide the best approximation. In equation (\ref{cosline}), $l$ is a binary variable indicating the range of $x$. When $l$=0, it implies that 
$-0.6\leq x \leq0$, and the cosine function will be approximated by the first line ($s_1x+m_1$)
as described in equations (\ref{xrange}) and (\ref{xmin}). When 
$l$=1, it indicates that $0\leq x \leq 0.6$, and the second line ($s_2x+m_2$)
will be used for approximation. The multiplication of binary and continuous variables in (\ref{cosline}) can be linearized using the big-M method. Figure \ref{fig. 2} illustrates the approximation of the cosine function using two linear segments. For the first segment, the slope $s$ is 0.24 and for the second segment $s$ is $-0.24$, with both having the same intercept $m$=1. This approximation yields a minimal error of 
3.58 \%, which is negligible. A similar approach can be applied to the sinus function. Figure \ref{fig. 3} shows the sinus function approximated by a single line with $s$ = 0.95 and $m$ = 0, achieving an error of 
0.94\%. %, which is the best approximation based on the defined process. 
This AC power flow linearization enhanced the model in \cite{baharvandi2019linearized}, offering improvements in both error reduction and the number of constraints.

\begin{figure}[h!]
	\centering
		\includegraphics[width=0.4\textwidth,height=0.10\textheight]{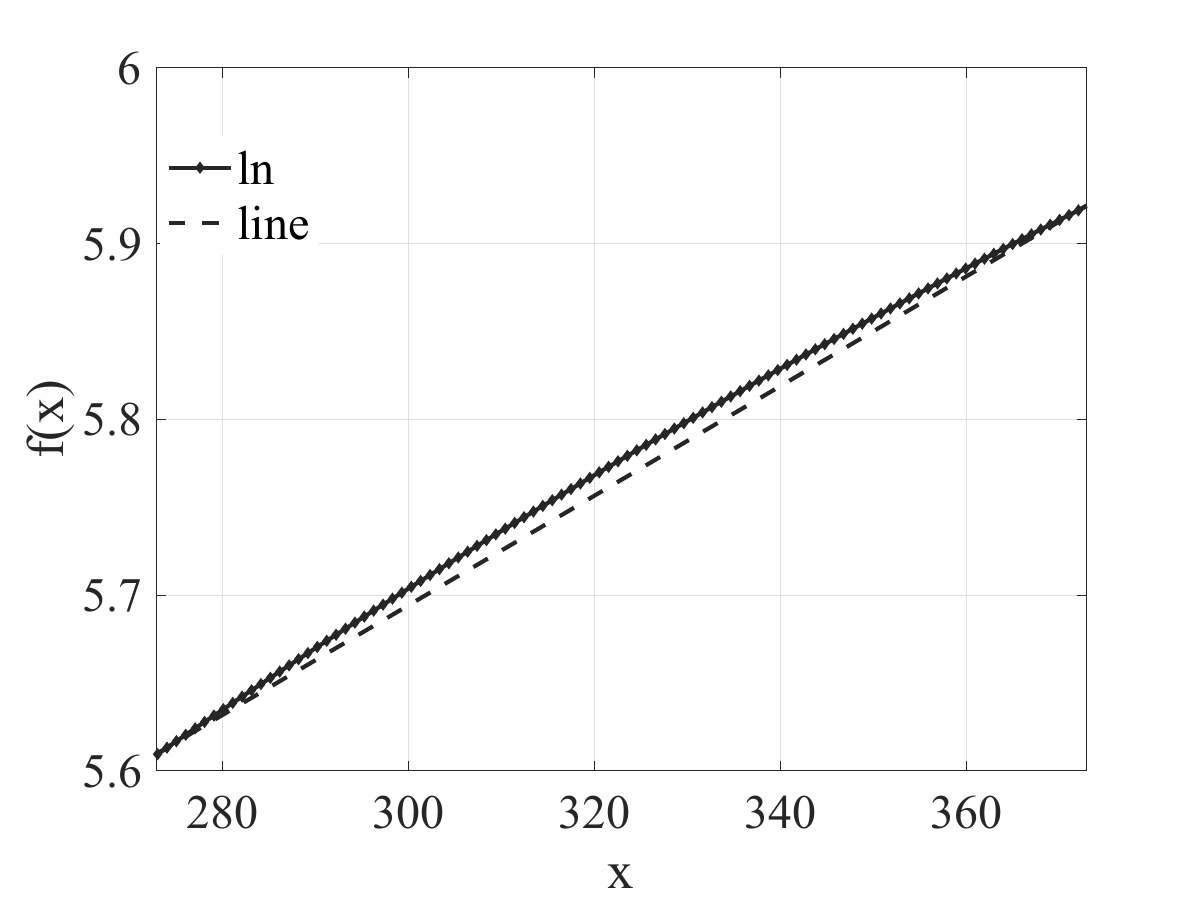}
			\caption{Approximation of $ln$ function to line}
	\label{fig. 4}
\end{figure}

The radiation losses equation in (\ref{radiation}) is expressed as a fourth-degree polynomial function. Given that line temperatures can vary significantly, though typically remaining below 373 K, a large number of linear segments would be needed to approximate this polynomial function accurately. To address this challenge, the following formulations can be applied:
\begin{align}
&ln(Q^{rad}_{c,d}(T_{c,d})+\epsilon K^{r}_{c,d} (T^{E}_{c,d})^{4})=ln(\epsilon K^{r}_{c,d} (T_{c,d})^{4})
\label{ln1}\\
&ln(Q^{rad}_{c,d}(T_{c,d})+\epsilon K^{r}_{c,d} (T^{E}_{c,d})^{4})=ln(\epsilon K^{r}_{c,d})+4ln(T_{c,d}))
\label{ln2}
\end{align}
Using the natural logarithm on both sides of (\ref{radiation}) results in (\ref{ln1}), which simplifies to (\ref{ln2}). The $\ln$ function is approximated using a single linear segment. Figure \ref{fig. 4} illustrates the linear approximation of $\ln(T_{c,d})$ with a line having a slope ($s$) of 0.00312 and a y-intercept ($m$) of 4.75824 over the temperature range of 273 K to 373 K. The maximum difference between the $\ln$ function and the line is 0.2 \%, highlighting the effectiveness of this linear approximation. Similarly, the left-hand side of equation (\ref{ln2}) is approximated with a line, which has a slope of 0.043 and a y-intercept of 1.917.


Additionally, to remove the absolute term in (\ref{active power formula}), the following constraints should be applied:
\begin{align}
&pf_{c,d}\leq d_{c,d} X
\label{active1}\\
&pf_{c,d}\geq -(1-d_{c,d}) X
\label{active2}\\
&V_b I_{c,d}=-pf_{c,d}+2\zeta_{c,d},
\label{active3}
\end{align}
where $d_{c,d}$ is a binary variable and $\zeta_{c,d}$ equals $d_{c,d} pf_{c,d}$ that can be linearized using the big-M method. If $d_{c,d}$ is zero, $pf_{c,d}$ will be negative according to (\ref{active1}), and $\zeta_{c,d}$ will be zero, so left side of (\ref{active3}) should be equal to the absolute of $pf_{c,d}$. If $d_{c,d}$ is 1, $pf_{c,d}$ will be positive based on (\ref{active2}), and $\zeta_{c,d}$ will be equal to $pf_{c,d}$. Then, the left side of (\ref{active3}) will satisfy the right side of (\ref{active power formula}). Equations (\ref{AC load flow}), (\ref{uncertain convection1}), and (\ref{uncertain convection2}) involve the multiplication of $u_c$ and continuous variables. The big-M method should be employed to linearize these constraints, 
\section{Simulation Results}
\label{Results}
%The problem with the uncertainty and DTLR is implemented
We conducted the experiments on the IEEE 6-bus and the IEEE 24-bus reliability test system (RTS).
All simulations have been carried out using the Gurobi solver in 
Python \cite{Python} on a personal computer with a core i7 processor and 8 GB RAM.

\subsection{IEEE 6-Bus Test System}
\label{IEEE 6-Bus Test System}
 The data related to the lines and units are extracted from \cite{roh2009market}. We consider a one year planning horizon that is divided into five 
subperiods. Every subperiod has a specific load factor, calculated as the ratio of the actual load and annual peak load. The load factors are 0.5, 0.65, 0.8, 0.9, and 1 for five periods, respectively. The weights of the loads located at buses 3, 4, and 5 are 0.4, 0.3, and 0.3, respectively. Each weight represents the proportion of the total peak load attributed to each bus. 
Three wind turbines are located in buses 3, 4, and 5. The average output power produced by each wind turbine is 20 MW. The PV systems are similar to the wind turbines in terms of their capacity, quantity, and placement. Details about the PV systems and wind turbines can be found in \cite{Output}. It is assumed that the EVs travel between cities and use charging stations installed on the highways that are related to buses 3, 4, and 5 in the test system. The forecast EV charging load at each bus is 10 MW.
All test systems are assessed in two categories. Case A is the IGTEP 
problem under uncertainty, without DTLR consideration. In case B, the DTLR is taken into account.

\begin{table}[ht] 
\centering
\caption{Simulation results without DTLR}
\begin{tabular}{|c|c|c|c|c|}
\hline
 \vtop{\hbox{\strut  Annual peak}\hbox{\strut demand (MW)}} & \vtop{\hbox{\strut  Added}\hbox{\strut lines}} &  \vtop{\hbox{\strut  Added}\hbox{\strut units}} &  \vtop{\hbox{\strut  Added}\hbox{\strut element }\hbox{\strut number}}& \vtop{\hbox{\strut  Objective}\hbox{\strut value ($\$10^7$)}}   \\
\hline
300 & $L_2$ & - & 1 &  3.635\\
\hline
350 & $L_4$ &  $G_2$,$G_8$ & 3 & 4.981\\
\hline
400 & \vtop{\hbox{\strut  $L_1$,$L_2$, }\hbox{\strut $L_6$}} & \vtop{\hbox{\strut  $G_2$,$G_7$, }\hbox{\strut $G_8$}}& 6 & 6.508\\
\hline
450 & $L_4$,$L_6$  & \vtop{\hbox{\strut  $G_2$,$G_4$, }\hbox{\strut $G_5$}}& 5 & 7.927\\
\hline
500 & \vtop{\hbox{\strut  $L_1$,$L_2$, }\hbox{\strut $L_3$,$L_6$}} & \vtop{\hbox{\strut  $U_1$,$G_2$, }\hbox{\strut $G_4$,$G_7$,}\hbox{\strut $G_8$}}& 9 & 9.439\\
\hline
550 & \vtop{\hbox{\strut  $L_1$,$L_2$, }\hbox{\strut $L_3$,$L_6$, } } & \vtop{\hbox{\strut  $U_1$,$U_5$, }\hbox{\strut $G_4$,$G_5$,}\hbox{\strut $G_8$}}& 9 & 10.922\\
\hline
600 & \vtop{\hbox{\strut  $L_1$,$L_2$, }\hbox{\strut $L_3$,$L_6$ } } & \vtop{\hbox{\strut  $U_1$,$U_5$, }\hbox{\strut $G_1$,$G_4$,}\hbox{\strut $G_5$,$G_7$}}& 10 & 12.458\\
\hline
650 & \vtop{\hbox{\strut  $L_1$,$L_2$, }\hbox{\strut $L_3$,$L_4$, } \hbox{\strut $L_6$,$L_7$}} & \vtop{\hbox{\strut  $U_1$,$U_4$, }\hbox{\strut $U_5$,$G_1$,}\hbox{\strut $G_2$,$G_4$,}\hbox{\strut $G_5$,$G_7$,} \hbox{\strut $G_8$}}& 15 & 14.096\\
\hline
700 & \vtop{\hbox{\strut  $L_1$,$L_2$, }\hbox{\strut $L_3$,$L_4$, } \hbox{\strut $L_6$,$L_7$}} & \vtop{\hbox{\strut  $U_1$,$U_4$, }\hbox{\strut $U_5$,$G_1$,}\hbox{\strut $G_3$,$G_4$,}\hbox{\strut $G_5$,$G_6$,} \hbox{\strut $G_8$}}& 15 & 15.876\\
\hline
750 & \vtop{\hbox{\strut  $L_1$,$L_2$, }\hbox{\strut $L_3$,$L_4$, } \hbox{\strut $L_5$,$L_6$,} \hbox{\strut $L_7$}} & \vtop{\hbox{\strut  $U_1$,$U_3$, }\hbox{\strut $U_4$,$U_5$,}\hbox{\strut $G_1$,$G_3$,}\hbox{\strut $G_4$,$G_5$,} \hbox{\strut $G_6$,$G_7$}}& 17 & 17.674\\
\hline
800 &- &-&0&Infeasible\\
\hline

\end{tabular}
\label{6bus without DTLR}
%\lable{Operation Cost}
\end{table}

\subsubsection{Case A: IGTEP problem without DTLR}
In this case, DTLR is not considered. %, so we use the formulation presented in section \ref{without DTLR}. 
Table \ref{6bus without DTLR} presents the simulation results for varying peak demand. The set of candidate lines is $\Upsilon_{L}^{\sf  N}$ = [$L_1$,$L_2$,...,$L_7$], and the set of candidate generators is $\Upsilon_{G}^{\sf  N}$ = [$U_1$,...,$U_5$,$G_1$,...,$G_8$]. The data for the candidate lines and generators are extracted from \cite{roh2009market}. As expected, increased demand necessitates more new lines and units to maintain the power balance. % between the generation and load. 
Furthermore, the total cost (i.e., the objective value) rises as the number of added lines and units increases. For instance, at a peak demand of 550 MW, 9 elements are added to the network, whereas for a 650 MW load, this number increases to 15. Correspondingly, the total cost increases from $10.922 \times 10^7$ to $14.096 \times 10^7$, reflecting the additional costs incurred by including new elements required to accommodate the higher load. Note that when the load reaches a sufficiently high level (e.g., 800 MW), the added generators and lines are insufficient to meet the demand, resulting in infeasibility.

\subsubsection{Case B: IGTEP problem with DTLR}
We examine the impact of DTLR on the optimal solution. The data for the model incorporating DTLR are provided in Tables \ref{line data1}--\ref{line data3}. The base apparent power and base voltage are set to 100 MVA and 132 kV, respectively. The simulation results are presented in Table \ref{6bus with DTLR}. We can observe that the total costs (i.e., the objective values) with DTLR are lower than those without DTLR. For example, when the annual peak load is 750 MW, the total cost in Case $A$ is $\$ 17.674\times 10^{7}$,   while in Case B, with DTLR, the total cost decreases to $\$ 17.077\times 10^{7}$.
Notably, incorporating DTLR allows for dynamic increases in the capacity of transmission lines, improving line utilization. This reduces the need to install new lines, as operators can leverage DTLR to adaptively enhance the capacity of existing lines. Consequently, the total cost and the number of newly added elements decrease. Additionally, DTLR enables the system to accommodate higher peak loads. Specifically, with DTLR, the IGTEP problem becomes infeasible at a peak load of 900 MW, whereas without DTLR (Case A), infeasibility occurs at an 800 MW peak load. These findings highlight the advantages of incorporating DTLR into the IGTEP problem, improving both cost efficiency and system capacity.

\begin{table}
\begin{center}

\caption{Data for lines}
\begin{tabular}{ |c|c|c|c|c|}

 \hline
 $R(T_c^{max})(\Omega)$&$G_c(P.U.)$ &$\beta_c(P.U.)$ & Length (Km)&$T_c^{max}$($^\circ$C)\\ 
 \hline
 10&1.024&4.099&50&100 \\ 
 \hline

\end{tabular}
\label{line data1}
\end{center}

\end{table}

\begin{table}
\begin{center}

\caption{Data for HBE}
\begin{tabular}{ |c|c|c|}

 \hline
   $Q_{c,d}^{s}(W/m)$ &$K_{c,d}^{r}(W/m-K^4)$&$\epsilon$ \\ 
 \hline
  14.08 &$2.5\times 10^{-9}$&0.75 \\ 
 \hline

\end{tabular}
\label{line data2}
\end{center}

\end{table}

\begin{table}
\centering
\caption{Properties of the conductor}
\begin{tabular}{ |c|c|c|c|c|} 
 \hline
 $\rho_c$ (kg/m\(^3\)) & $v_{c,d}$ (m/s) & \parbox{1cm}{$\nu_c$ \\ \strut (kg/m·s)} & $D_c$ (m) & \parbox{1cm}{$\gamma^{con}_c$ \\ \strut (W/m·K)}\\ 
 \hline
 1.293 & 2.23 & $1.81 \times 10^{-5}$ & 0.035 & 0.028 \\ 
 \hline
\end{tabular}
\label{line data3}
\end{table}

\begin{table}[ht] 
\centering
\caption{Simulation results with DTLR}
\begin{tabular}{|c|c|c|c|c|}
\hline
 \vtop{\hbox{\strut  Annual peak}\hbox{\strut demand (MW)}} & \vtop{\hbox{\strut  Added}\hbox{\strut lines}} &  \vtop{\hbox{\strut  Added}\hbox{\strut units}} &  \vtop{\hbox{\strut  Added}\hbox{\strut element }\hbox{\strut number}}& \vtop{\hbox{\strut  Objective}\hbox{\strut function  ($\$10^7$)}}   \\
\hline
300 & - & - & 0 & 3.574\\
\hline
350 & - & $U_4$& 1 & 4.882\\
\hline
400 & - &  $U_1$ & 1 & 6.353\\
\hline
450 & - & \vtop{\hbox{\strut  $U_1$,$U_4$, }\hbox{\strut $G_8$} }& 3 & 7.697\\
\hline
500 &- & \vtop{\hbox{\strut  $U_1$,$U_4$, }\hbox{\strut $G_2$,$G_7$,}\hbox{\strut $G_8$}}& 5 & 9.192\\
\hline
550 &- & \vtop{\hbox{\strut  $U_1$,$U_4$, }\hbox{\strut $G_4$,$G_5$,}\hbox{\strut $G_8$}}& 5 & 10.667\\
\hline
600 & - & \vtop{\hbox{\strut  $U_1$,$U_4$, }\hbox{\strut $G_2$,$G_4$,}\hbox{\strut $G_5$,$G_7$,}\hbox{\strut $G_8$}}& 7 & 12.168\\
\hline
650 & - & \vtop{\hbox{\strut  $U_1$,$U_4$, }\hbox{\strut $U_5$,$G_1$,}\hbox{\strut $G_2$,$G_4$,}\hbox{\strut $G_5$,$G_7$,} \hbox{\strut $G_8$}}& 9 & 13.778\\
\hline
700 &$L_6$ & \vtop{\hbox{\strut  $U_1$,$U_4$, }\hbox{\strut $U_5$,$G_1$,}\hbox{\strut $G_4$,$G_5$,}\hbox{\strut $G_6$,$G_7$,} \hbox{\strut $G_8$}}& 10 & 15.430\\
\hline
750 &$L_3$ &\vtop{\hbox{\strut  $U_1$,$U_2$, }\hbox{\strut $U_4$,$U_5$,}\hbox{\strut $G_2$,$G_4$,}\hbox{\strut $G_5$,$G_6$,} \hbox{\strut $G_7$,$G_8$}}& 11 & 17.077\\
\hline
800 &$L_3$,$L_7$ &\vtop{\hbox{\strut  $U_1$,$U_2$, }\hbox{\strut $U_3$,$U_4$,}\hbox{\strut $U_5$,$G_1$,}\hbox{\strut $G_4$,$G_5$,} \hbox{\strut $G_6$,$G_7$,}\hbox{\strut $G_8$}}& 13 & 18.795\\
\hline
850 &$L_3$ &\vtop{\hbox{\strut  $U_1$,$U_2$, }\hbox{\strut $U_3$,$U_4$,}\hbox{\strut $U_5$,$G_1$,}\hbox{\strut $G_2$,$G_3$,} \hbox{\strut $G_4$,$G_5$}\hbox{\strut $G_6$,$G_7$,}\hbox{\strut $G_8$}}& 14 &  20.485\\
\hline
900 &- & -& 0 & Infeasible\\
\hline

\end{tabular}
\label{6bus with DTLR}
%\lable{Operation Cost}
\end{table}

 \subsection{IEEE 24-Bus RTS}
\label{IEEE 24-Bus RTS}
The network comprises 32 existing generators and 34 transmission lines. Details about this network are available in \cite{grigg1999reliability} and Table \ref{24bus line data}. The load factor and duration of subperiods are identical to those used in the IEEE 6-bus test system. 
The total output power from the PVs and wind turbines is 180 MW, with each wind/solar generator contributing 30 MW. These RESs are situated at buses 1, 6, 9, 13, 16, and 20. Additionally, the projected power consumption by electric vehicles (EVs) in this test system is 30 MW, located at the same buses as the PVs.
Simulations were conducted for both scenarios, with and without DTLR, and the results are presented in Tables \ref{24bus without DTLR} and \ref{24bus with DTLR}. In this study, all generating units, except for hydro units, were evaluated as candidate generators. Detailed data for the candidate transmission lines are provided in \cite{aghaei2014generation}. The set of candidate lines is $\Upsilon_{L}^{\sf  N}$ = [$L_1$,$L_2$,...,$L_{10}$], and the set of candidate generators is $\Upsilon_{G}^{\sf  N}$ = [$U_1$,$U_2$,...,$U_{26}$]. As anticipated, the objective value in Case B (with DTLR) is lower than in Case A (without DTLR), confirming the positive effect of DTLR on cost reduction. For instance, under an annual peak load of 4100 MW, the total cost is $\$ 2.037 \times 10^7$ in the problem without DTLR, while this amount decreases to $\$ 2.004 \times 10^7$ in the problem with DTLR. Furthermore, DTLR allows the system to accommodate higher peak loads. For example, without DTLR, the system becomes infeasible at a demand of 4400 MW, whereas with DTLR, infeasibility occurs at 4500 MW. In this case study, it was observed that the current transmission lines possess adequate capacity to meet the demand. Therefore, there is no requirement for the addition of new lines at this time. Table \ref{computational time} compares the results for the linear and non-linear DTLR models applied to the IEEE 24-bus RTS with a 4200 MW load. The objective values for both problems are similar; however, there is a significant difference in computation time. The linear model requires 17.31 seconds to solve, whereas the non-linear model takes 1532.03 seconds. This demonstrates the efficiency of the linearized approach in significantly reducing computational time.

\begin{table}
\begin{center}
\caption{Line data}
\begin{tabular}{ |c|c|c|c|c|}

 \hline
 $R(T_c^{max})(\Omega)$&$G_c(P.U.)$ &$\beta_c(P.U.)$ & Length (Km)&$T_c^{max}$(C)\\ 
 \hline
 1.76&1.024&10&125&100 \\ 
 \hline
 
\end{tabular}
\label{24bus line data}
\end{center}
\end{table}

\begin{table}[ht] 
\centering
\caption{Simulation results without DTLR}
\begin{tabular}{|c|c|c|c|c|}
\hline
 \vtop{\hbox{\strut  Annual peak}\hbox{\strut demand (MW)}} & \vtop{\hbox{\strut  Added}\hbox{\strut lines}} &  \vtop{\hbox{\strut  Added}\hbox{\strut units}} &  \vtop{\hbox{\strut  Added}\hbox{\strut element }\hbox{\strut number}}& \vtop{\hbox{\strut  Objective}\hbox{\strut value (\$ $10^7$)}}   \\
\hline
3500 & - & $U_{22}$,$U_{23}$ & 2 & 1.427\\
\hline
3800 & - & \vtop{\hbox{\strut  $U_{13}$,$U_{22}$, }\hbox{\strut $U_{23}$}} & 3 & 1.685\\
\hline
4100 & - & \vtop{\hbox{\strut  $U_1$,$U_2$, }\hbox{\strut $U_5$,$U_6$,} \hbox{\strut $U_{12}$,$U_{14}$,} \hbox{\strut $U_{16}$,$U_{17}$,} \hbox{\strut  $U_{18}$,$U_{19}$,} \hbox{\strut $U_{22}$,$U_{23}$}}& 12 & 2.037\\
\hline
4400 &- & -& 0 & Infeasible\\
\hline

\end{tabular}
\label{24bus without DTLR}
%\lable{Operation Cost}
\end{table}

\begin{table}[ht] 
\centering
\caption{Simulation results with DTLR}
\begin{tabular}{|c|c|c|c|c|}
\hline
 \vtop{\hbox{\strut  Annual peak}\hbox{\strut demand (MW)}} & \vtop{\hbox{\strut  Added}\hbox{\strut lines}} &  \vtop{\hbox{\strut  Added}\hbox{\strut units}} &  \vtop{\hbox{\strut  Added}\hbox{\strut element }\hbox{\strut number}}& \vtop{\hbox{\strut  Objective}\hbox{\strut value ($\$10^7$)}}   \\
\hline
3500 & - & $U_{22}$,$U_{23}$ & 2 & 1.400\\
\hline
3800 & - & \vtop{\hbox{\strut  $U_{14}$,$U_{22}$, }\hbox{\strut $U_{23}$}} & 3 & 1.664\\
\hline
4100 & - & \vtop{\hbox{\strut  $U_1$,$U_2$, }\hbox{\strut $U_5$,$U_6$,} \hbox{\strut $U_{12}$,$U_{13}$,} \hbox{\strut $U_{15}$,$U_{16}$,} \hbox{\strut  $U_{17}$,$U_{22}$,} \hbox{\strut $U_{23}$}}& 11 & 2.004\\
\hline
4400 & - & \vtop{\hbox{\strut  $U_{12}$,$U_{13}$, }\hbox{\strut $U_{14}$,$U_{20}$,} \hbox{\strut $U_{21}$,$U_{22}$,} \hbox{\strut $U_{23}$}}& 7 & 2.429\\
\hline
4500 & - & -& 0 & Infeasible\\
\hline

\end{tabular}
\label{24bus with DTLR}
%\lable{Operation Cost}
\end{table}

\begin{table}[h!]
\begin{center}
\caption{Comparison of linear and non-linear problem }
\begin{tabular}{ |c|c|c|} 
        
 \hline
& Objective function ($\$10^7$)& Computational time (s)\\ 
 \hline
 Linear&2.145&17.31 \\ 
 \hline
 Non-linear&2.111&1532.03 \\ 
 \hline
 
\end{tabular}
\label{computational time}
\end{center}
\end{table}

\section{Conclusion}
\label{Conclusion}
This paper proposed a new model to address the IGTEP problem, incorporating RES, EV charging, and DTLR. Through simulations on both the IEEE 6-bus and IEEE 24-bus test systems, we have demonstrated that the integration of DTLR improves the efficiency and resilience of the transmission network by dynamically adjusting line capacities based on real-time environmental conditions. This approach reduces the need for new infrastructure, thus lowering total investment and operational costs while ensuring the system can handle higher peak loads. Furthermore, we showed that incorporating DTLR delays system infeasibility at higher demand levels, enabling the grid to accommodate more load. The comparison between linear and non-linear DTLR models also highlighted the computational advantages of linearization, significantly reducing solution times without compromising on performance.

% The IGTEP problem, incorporating DTLR along with PVs, wind turbines, and EVs faces uncertainties addressed by a method combining RO and SP. This approach effectively manages the diverse uncertainties involved. DTLR provides operators with improved insights into line capacity, which is particularly valuable over long planning periods where temperature variations become significant. As a result, the inclusion of DTLR leads to a reduction in the objective function value compared to scenarios without DTLR, enhancing network efficiency.

% The expansion planning, SP/RO method, AC load flow, and HBE introduce nonlinearities that are linearized by replacing them with inequalities or affine equations. Simulation results demonstrate that the proposed methodology effectively addresses the uncertainties in expansion planning and highlights the impact of DTLR on test systems. Additionally, the linearization of the problem facilitates simpler computations, proving especially beneficial for large-scale real-world networks.

\bibliographystyle{IEEEtran}
\bibliography{reference}

\end{document}